\documentclass[a4paper,11pt]{article}

\usepackage[english]{babel}
\usepackage[utf8x]{inputenc}
\usepackage[T1]{fontenc}
\usepackage{authblk}
\usepackage{float}
\usepackage{booktabs}
\usepackage[font=small,labelfont=bf,margin=1.5cm]{caption}

\usepackage[a4paper,top=2cm,bottom=2cm,left=2.5cm,right=2.5cm,marginparwidth=1.75cm]{geometry}

\usepackage{amsmath}
\usepackage{graphicx}
\usepackage[colorinlistoftodos]{todonotes}
\usepackage[colorlinks=true, allcolors=blue]{hyperref}

\title{\bf Distance entropy cartography characterises centrality in complex networks}

\author[1]{Massimo Stella}
\author[2]{Manlio De Domenico}
\affil[1]{Fondazione Bruno Kessler; massimo.stella@inbox.com}
\affil[2]{Fondazione Bruno Kessler; mdedomenico@fbk.eu}

\begin{document}
\maketitle

\abstract{We introduce distance entropy as a measure of homogeneity in the distribution of path lengths between a given node and its neighbours in a complex network. Distance entropy defines a new centrality measure whose properties are investigated for a variety of synthetic network models. By coupling distance entropy information with closeness centrality, we introduce a network cartography which allows one to reduce the degeneracy of ranking based on closeness alone. We apply this methodology to the empirical multiplex lexical network encoding the linguistic relationships known to English speaking toddlers. We show that the distance entropy cartography better predicts how children learn words compared to closeness centrality. Our results highlight the importance of distance entropy for gaining insights from distance patterns in complex networks.}

\vspace{0.5cm}


\section{Introduction}

Defining the centrality of nodes in complex networks is an important question for determining the role of individual agents in a variety of dynamical processes such as information flow and influence maximisation \cite{borgatti2005centrality,morone2015influence,brede2018competitive}, network growth \cite{albert2002statistical} and resilience to cascade failures \cite{buldyrev2010catastrophic,de2017modeling}. 

In many of these processes, centrality can be defined by means of distance, i.e. the minimum number of links separating any two nodes. Overwhelming evidence from real-world network analysis has shown how distance among nodes is an important indicator for the evolution of a given process: in general, information flows at slower rates between nodes at greater distance in social networks \cite{borgatti2005centrality}; the recollection of words at greater distance in semantic networks in memory tasks is slower \cite{collins1975spreading}; smaller network distance between oscillators facilitates synchronization \cite{brede2018competitive} and consensus \cite{de2017diffusion}; there is a higher turnover rate of animal and plant species among closer sites in river networks \cite{rouquette2013species}. 

One type of network distance can be used for quantifying node's centrality in spreading processes where the flow follows shortest paths and such centrality is called closeness \cite{bavelas1950communication,borgatti2005centrality}. Closeness centrality quantifies the average distance of all the network paths leading to a given node. In undirected, unweighted networks, closeness centrality $c_i$ of node $i$ ($i=1,2,...,N$) is defined as \cite{bavelas1950communication}:

\begin{equation}
c_{i}=\frac{C_{i}}{\sum\limits_{j\neq i}d_{ij}},
\end{equation}

where $C_i$ is the number of nodes in the same connected component of $i$ and $d_{ij}$ is the network distance between nodes $i$ and $j$. It is worth nothing that $C_i=N$ for networks with a single connected component. It has been shown that this estimator is ill-defined in the case of disconnected networks \cite{marchiori2000harmony}: however, in the following we will deal only with connected networks and we can safely use the above definition of closeness centrality.

\section{Introducing the distance entropy}

Closeness centrality represents the inverse of the mean value of the distribution of path lengths from a given node to the rest of the system. Hence, closeness alone offers no information about the spread of the distribution of network distances. To account for this spread we introduce the \textit{distance entropy} $h(i)$, defined as the information entropy of the set $\textbf{d}^{(i)}\equiv(d_{i1},...,d_{ij},...,d_{iN})$ of distances between node $i$ and any other node $j$ in the system, here assumed to be of size $N$ ($1 \leq j \leq N$). Let us denote with $M_i$ the maximum distance $M_i=\text{max}_j d_{ij}$ and with $m_i$ the minimum distance $m_i=\text{min}_j d_{ij}$. Let us denote by $p_k^{(i)}=P(d_{ij}=k)$ with $m_i \leq k \leq M_i$, the probability that the generic entry $d_{ij}$ is equal to $k$. We exclude the distance of a node from itself. In formulas, we define distance entropy as:

\begin{equation}
h(i)=-\frac{1}{\text{log}(M_i-m_i)} \sum_{k=1}^{M_i-m_i} p_{k}^{(i)}\text{log}p_{k}^{(i)}.
\end{equation}

With this definition $h(i)$ ranges between 0 and 1. We can interpret the meaning of these extremal values in terms of network centrality by considering the entropy distance of regular graphs.

\subsection{Distance entropy in regular graphs}

Let us consider a complete graph of $N$ nodes, $\mathcal{K}_N$. In this graph, node $i$ is at distance 1 from all the other nodes, so then $\textbf{d}^{(i)}=(1,...,1)$ is a set of $N-1$ entries, all equal to 1. It is straightforward to verify that, in this case, the information entropy for node $i$ is $h(i)=0$. The same analysis holds for all the other $N-1$ nodes in the complete graph. Hence, all the nodes in a complete graph have distance entropy $h=0$. More in general, distance entropy is 0 for all nodes adjacent to all other nodes in any given network.

In star graphs there is a central node connected to all other peripheral nodes and no other links are present. The result for nodes in complete graphs holds also for the star centre. Consequently, the centres of star graphs have distance entropy $h=0$. Hence, we can interpret distance entropy as a measure of regularity of the distribution of path lengths between a given node and its neighbours, with $h=0$ representing the case of maximum homogeneity in the path length distribution. Since it is not possible for a node to be at distance $d \geq 2$ from all other nodes simultaneously in a connected network, then $h=0$ identifies nodes adjacent to all other nodes.

In a ring graph where $N>3$ nodes have only two neighbours, then every node has the same set of distances to the other $N-1$ nodes and the possible distances are ${1,2,...,\left\lfloor N/2\right\rfloor}$, where $\left\lfloor . \right\rfloor$ is the floor function. If $N$ is odd, then it is easy to check that $p_{k}=2/(N-1)$. Consequently, the entropy of any node $i$ in a ring graph with an odd number of nodes is:

\begin{equation}
h(i)=-\frac{\left\lfloor N/2\right\rfloor }{\text{log}\left\lfloor N/2\right\rfloor }(\frac{2}{N-1})\text{log}(\frac{2}{N-1})=-\frac{1}{\text{log}[(N-1)/2]}\text{log}(\frac{2}{N-1}).
\end{equation}

As an example, when $N=5$ then all nodes have maximum distance entropy $h=1$, which corresponds to the case of minimum homogeneity of path lengths, i.e. paths between connected nodes have lengths that are distributed uniformly across all possible distances.

When $N$ is even, then $p_{k}=2/(N-1)$ except for $k=N/2$, for which $p_{N/2}=1/(N-1)$. Then the formula for the distance entropy of a node becomes:

\begin{equation}
h(i)=-\frac{1}{\text{log}(N/2)}\left[ \frac{N-2}{N-1} \text{log}(\frac{2}{N-1})+\frac{1}{N-1}\text{log}(\frac{1}{N-1})\right].
\end{equation}

Notice that ring graphs are lattice graphs for which the coordination number $z=2$ (i.e. every node is connected only to two other nodes). While the analytical results for ring graphs can be extended also to cases for $z>2$, when $z$ approaches $N-1$ the lattice becomes a complete graph and hence $h\rightarrow 0$ for every node. When $z \ll N-1$, instead, results similar to the ring lattices hold and the lattice nodes are expected to have values of distance entropy close to 1. Rather than considering other regular structures we now focus on characterising patterns of distance entropy in network models frequently used in the relevant literature.

\section{Distance entropy and network models}

We consider three main network models usually considered in the literature: Erdos-Renyi (ER) random graphs \cite{erds1960evolution}, Watts-Strogatz small-world (SW) networks \cite{watts1998collective}, and Barabasi-Albert (BA) preferential attachment networks \cite{albert2002statistical}. For each network model, we are interested in characterising trends of the average distance entropy depending on the model parameters. In one case, we discover the presence of a tipping point for distance entropy in ER random graphs and relate it numerically to the appearance of short-cuts for increasing edge densities.

\subsection{Homogeneous random graphs}

\begin{figure}[H]
\centering
\includegraphics[width=7cm]{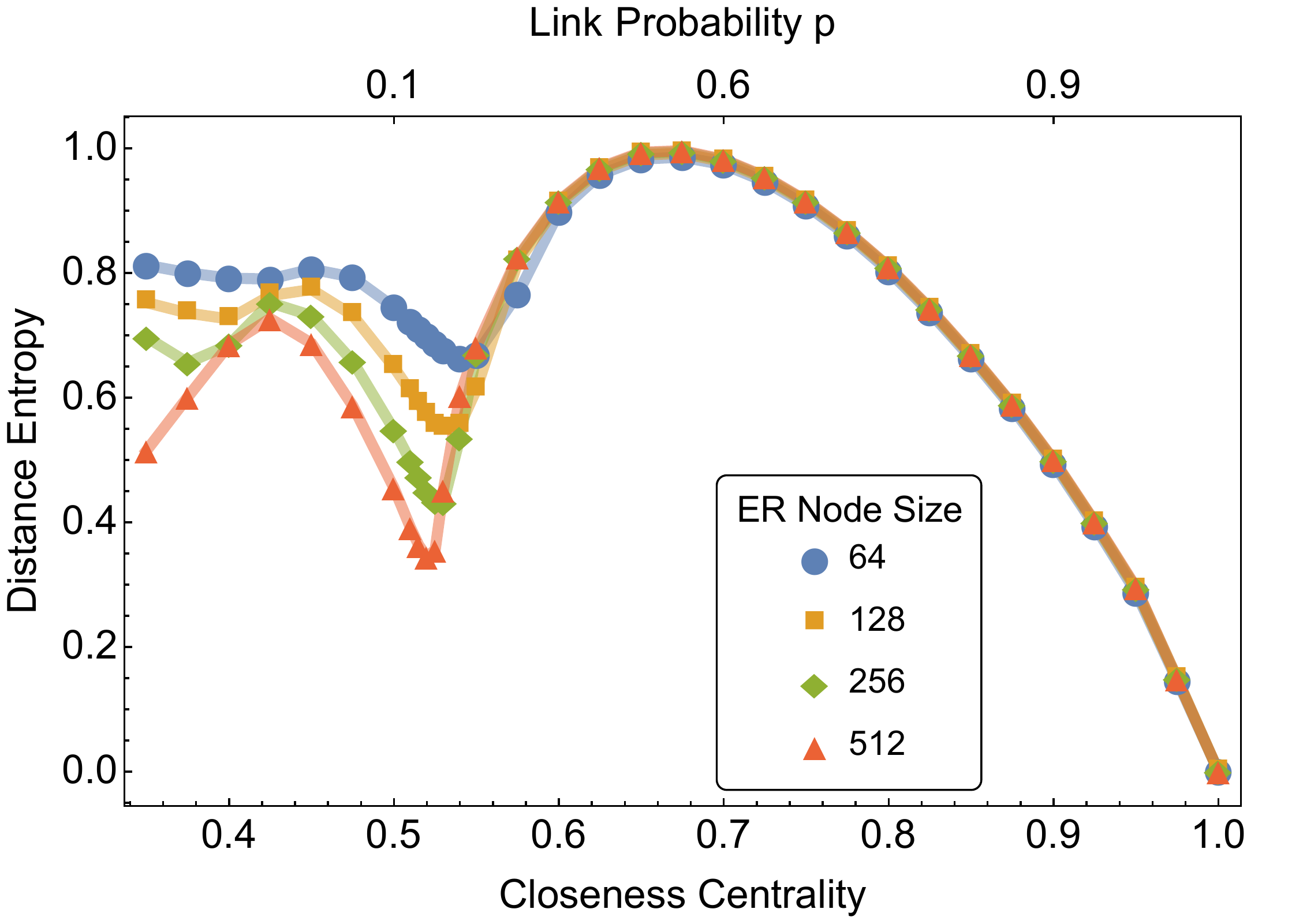}
\includegraphics[width=6.8cm]{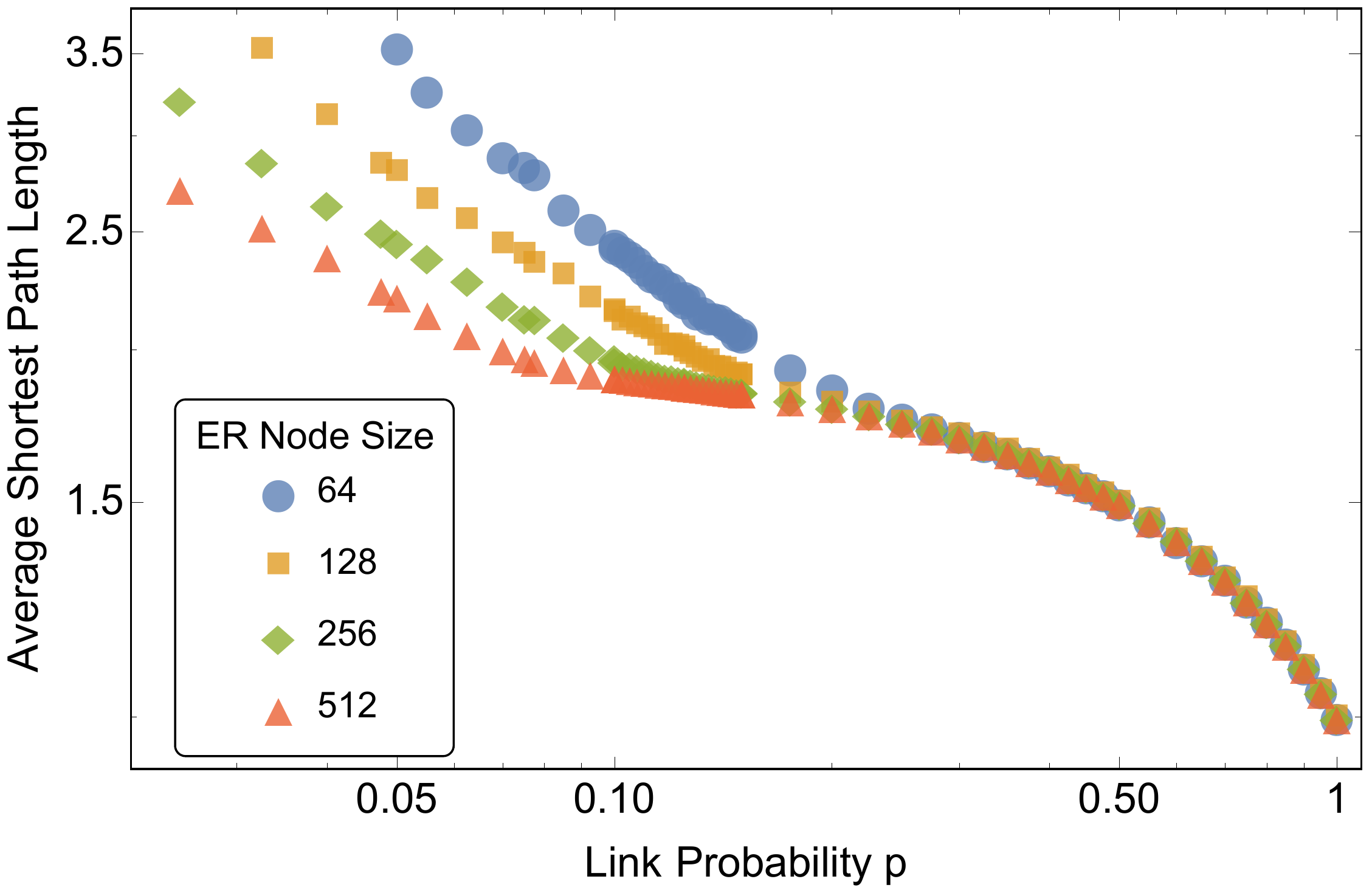}
\caption{\textbf{Left:} Mean closeness centrality and distance entropy of ER random graphs of different node sizes and different link probabilities. Mean values are averaged for all nodes in a graph and across 100 different graph realisations. Link probabilities are relative to all node sizes. All different ensembles converge to the same pattern of distance entropy roughly above rewiring probability $p=0.6$. \textbf{Right:} On a log-log scale, average shortest path length of ER random graphs for different link probabilities. The average distance decreases with increasing rewiring probability $p$ and tipping points are evident around $p \approx 0.1$, after which the average shortest path length decreases with a slower rate.}
\label{fig:random}
\end{figure}

Figure \ref{fig:random} (left) plots the mean closeness centrality and distance entropy of ER random graphs of different size $N$ and different link probability $p$. As expected, the addition of links makes random graphs closer to complete graphs, increasing the average closeness centrality when $p$ rises. Instead, numerical results on the distance entropy indicate that the entropy of path lengths in ER random graphs is not monotonic: we identify a tipping point, depending on system's size, in the average entropy that, asymptotically, converges to $p^{*} \approx 0.1$, which is well above the critical values for the emergence of the largest connected component $p_c=1/N$ and for connectedness $p_c=\text{log}N/N$. A thorough characterization of this tipping point, identifying a structural change of paths in ER random graphs, is out of the scope of the present work. It is interesting to notice that all the simulated ensembles converge towards the same pattern of distance entropy and closeness for increasing values of $p>p^*$, independently on their size.
Before the tipping point $p<p^*$, short-cuts (i.e. paths with length 1) start appearing in the networks when $p$ approaches $p^*$ from the left, as noticeable also from the trend of average shortest path length on $p$ (see Figure \ref{fig:random} (right)). Once created, short-cuts make nodes closer with each other, path lengths of shorter length start appearing with higher frequency and hence the average distance entropy reduces. However, the densification due to increases in $p$ happens at random, hence not all possible short-cuts are created in the network when $p$ is slightly higher than $p^*$. The random occurrence of edges will give rise to shortest paths having a homogeneous length distribution, corresponding to increases in the values of $h$ up to values close to 1. When even more links are added, the random graphs get closer to a complete graph, for which $h=0$, and hence the average distance entropy decreases. 

\subsection{Small-world networks}

\begin{figure}[H]
\centering
\includegraphics[width=9cm]{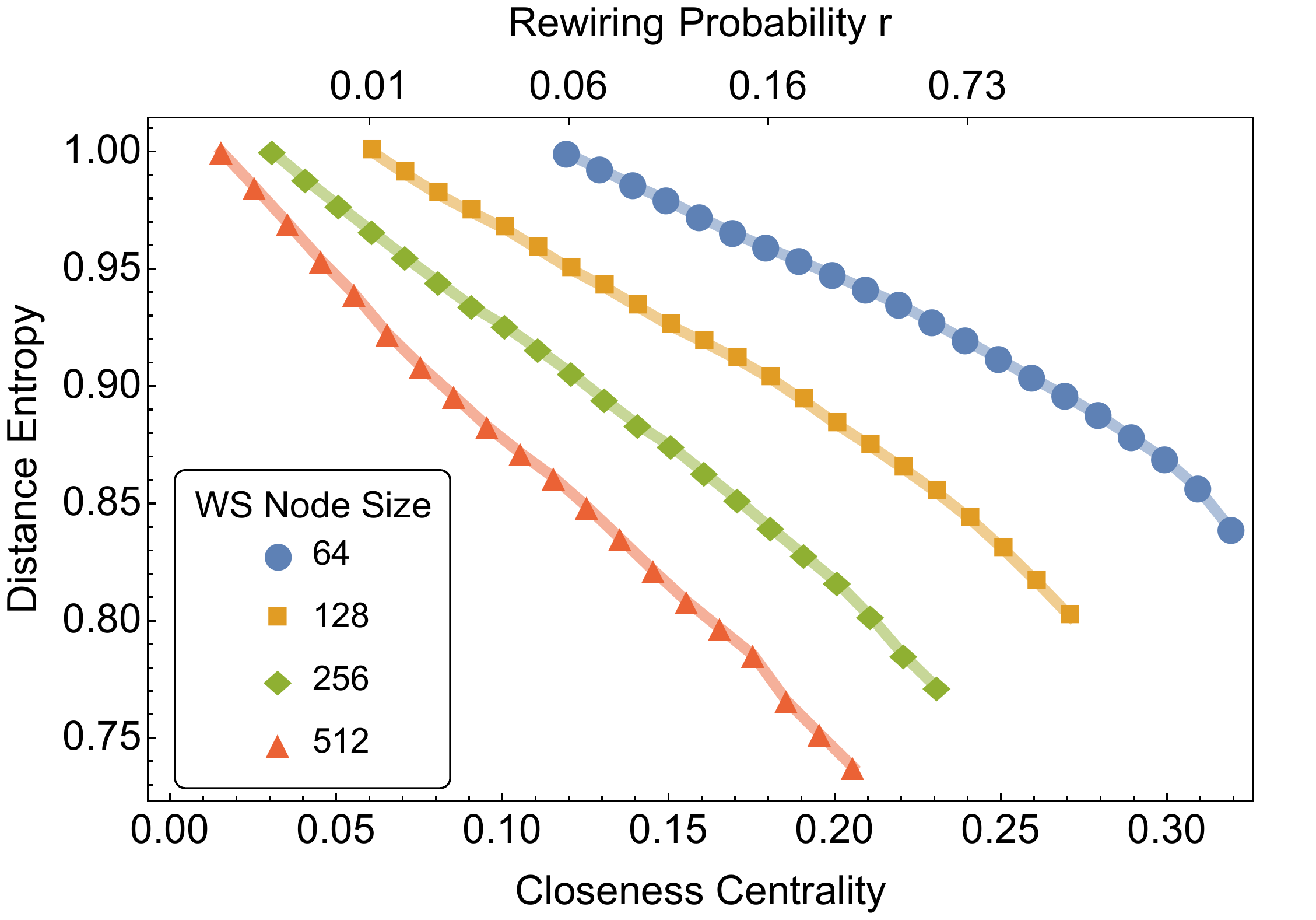}
\caption{Mean closeness centrality and distance entropy of SW networks of different node sizes and different rewiring probabilities. The rewiring probabilities plotted above are relative only to the case with $N=256$ and are provided only as a guideline. Mean values are averaged for all nodes in a graph and across 100 different independent realizations.}
\label{fig:smallworld}
\end{figure}

Figure \ref{fig:smallworld} reports the patterns of mean closeness centrality and distance entropy for small-world networks of different sizes, for coordination number $z=4$ and at different rewiring probabilities $r$. When $r=0$, a small-world network is a lattice with coordination number $z$. For intermediate values $0<r<1$ a fraction $r$ of the links for each node is rewired uniformly at random. For $r=1$ a small-world network is equivalent to a random graph with $p=z/N$ \cite{watts1998collective}.

Figure \ref{fig:smallworld} shows that the rewiring probability has a monotonic effect on the mean distance entropy, which decreases from values close to $1$ ($r=0$). This is expected, as the rewiring is increasingly destroying the order of the lattice structure, thus reducing $h$. Notice that the minimum value of distance entropy, reached when $r=1$, is the same mean distance entropy of a random graph with $p=z/N$. The numerical results in Fig. \ref{fig:smallworld} indicate that distance entropy does not detect the so-called small-world regime, i.e. a region determined by intermediate values of $r$ for which small-world display an average short path length close to $\text{log}N$ and values of clustering coefficient comparatively higher than those of a random graph. In fact, no tipping point relative to this phase transition are found in the numerical simulations, independently on the considered network size. We attribute this finding to the fact that distance entropy can only highlight deviations from homogeneous distribution of path lengths and it cannot provide information about either the assortative mixing or the clustering of nodes, which are both network features that have to be measured in order to characterise the small-world property.

\subsection{Barabasi-Albert networks}

\begin{figure}[H]
\centering
\includegraphics[width=9cm]{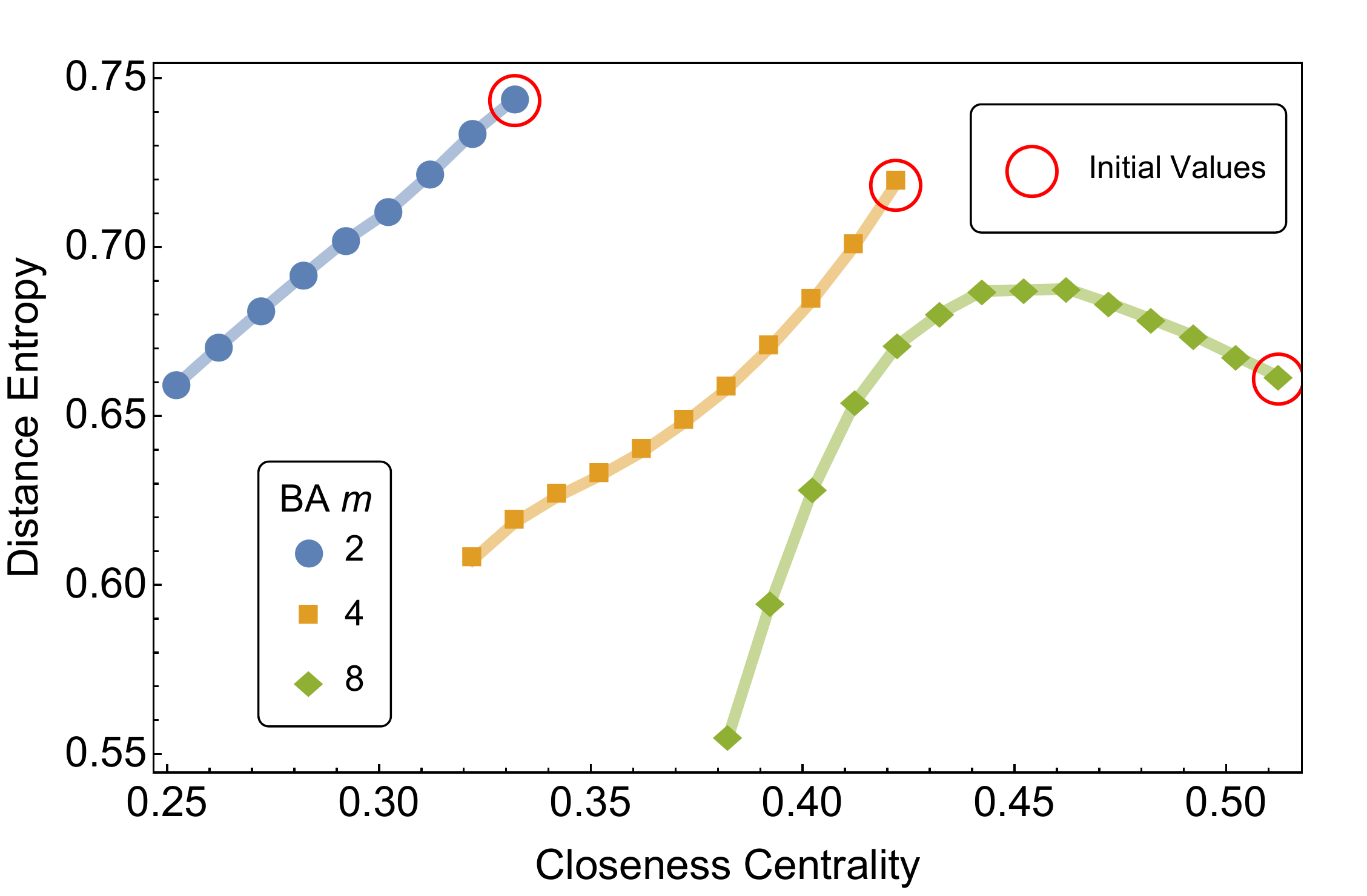}
\caption{Mean closeness centrality and distance entropy of \textit{growing} BA networks for different values of the link growth rate $m$. Initial values are relative to networks with $100$ nodes and are highlighted in red. Growing networks are measured once every 100 nodes are added. Mean values are averaged for all nodes in a graph and across 100 independent realizations.}
\label{fig:barabasi}
\end{figure}

Figure \ref{fig:barabasi} reports on the mean closeness centrality and distance entropy of growing BA network models. Network growth follows a preferential attachment process where one node and $m$ links are added at every time step \cite{albert2002statistical}. 

Since it is already known that the average network distance $l$ grows as $\text{log}N/\text{log log}N$ in growing BA networks with $N$ nodes, then it is excepted for closeness centrality to decrease with the growing number of nodes. When the number $m$ of links added at each time step is smaller than $8$ a monotonic decrease in distance entropy is registered during network growth. Instead, when $m=8$, smaller BA networks display a peak of distance entropy for intermediate sizes ($N\approx 600$). Entropy $h$ decreases at later steps when more nodes and links are added. We link this decrease in distance entropy with the emergence of hubs due to preferential attachment in larger BA networks. Hub nodes tend to have distance 1 from a significant fraction of nodes in the network, thus lowering considerably the average distance entropy. 

\section{Cartography based on distance entropy and closeness centrality}

In the previous sections we focused on characterising the mean distance entropy of networks by considering different models. We now focus on the structural patterns of individual nodes that emerge by considering closeness centrality and distance entropy together in one given network. We consider distance entropy as an estimator of the variation of distances of a given node from all its neighbours, thus providing additional information compared to considering closeness centrality only (which reports only on the mean distance of a node from the other nodes). 

We combine information from both closeness and distance entropy of a node by introducing the concept of \textit{distance entropy cartography}. We draw inspiration from the concept of cartography introduced by Guimer\'{a} and Amaral for characterising the role played by individual nodes in community structure \cite{guimera2005functional}, a concept later generalised to the participation of nodes on multiplex structures \cite{battiston2014structural}. Network cartographies are useful for visualising the map of topological patterns that nodes have in a given network structure.

\begin{figure}[H]
\centering
\includegraphics[width=10cm]{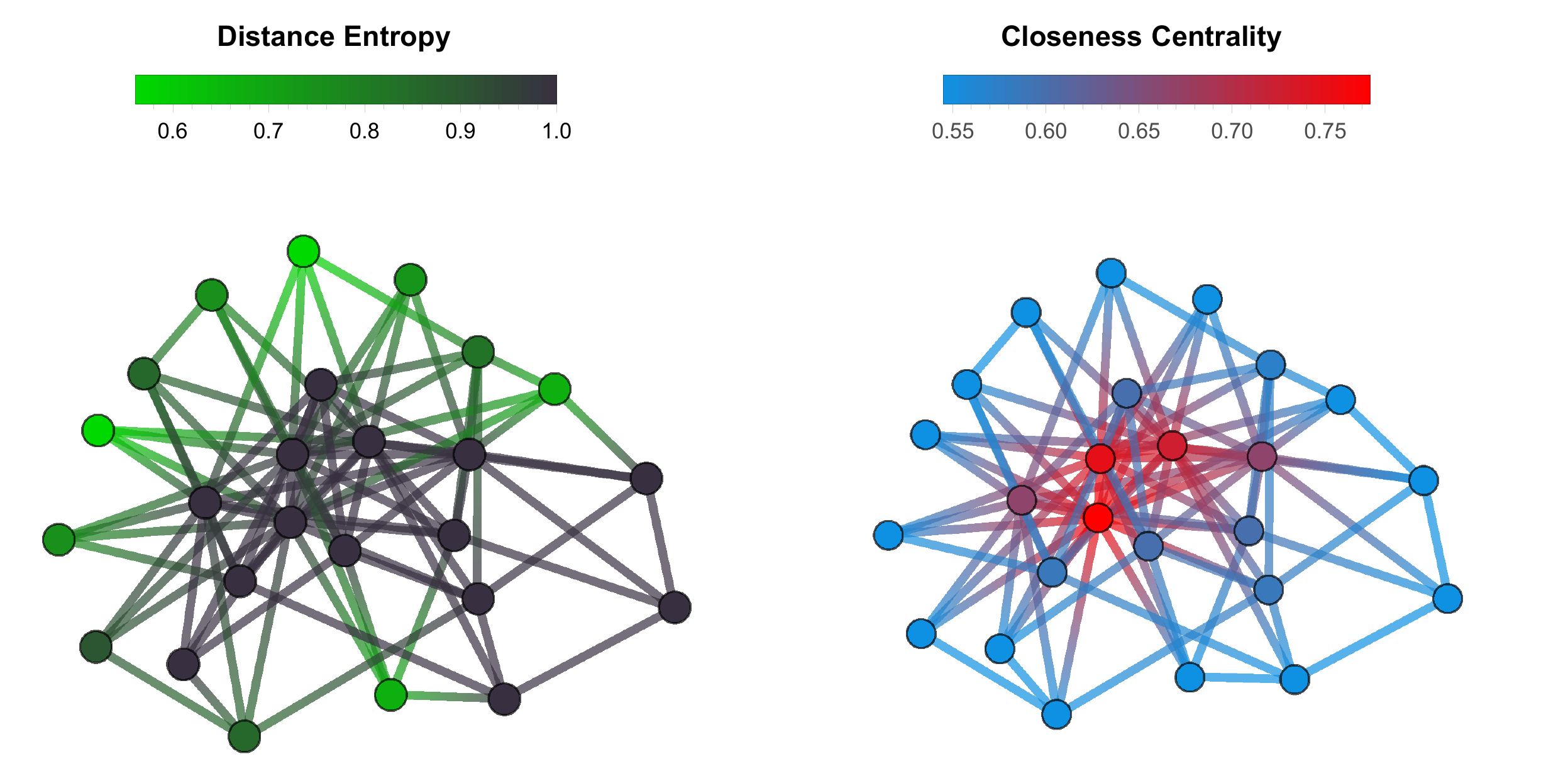}
\includegraphics[width=5cm]{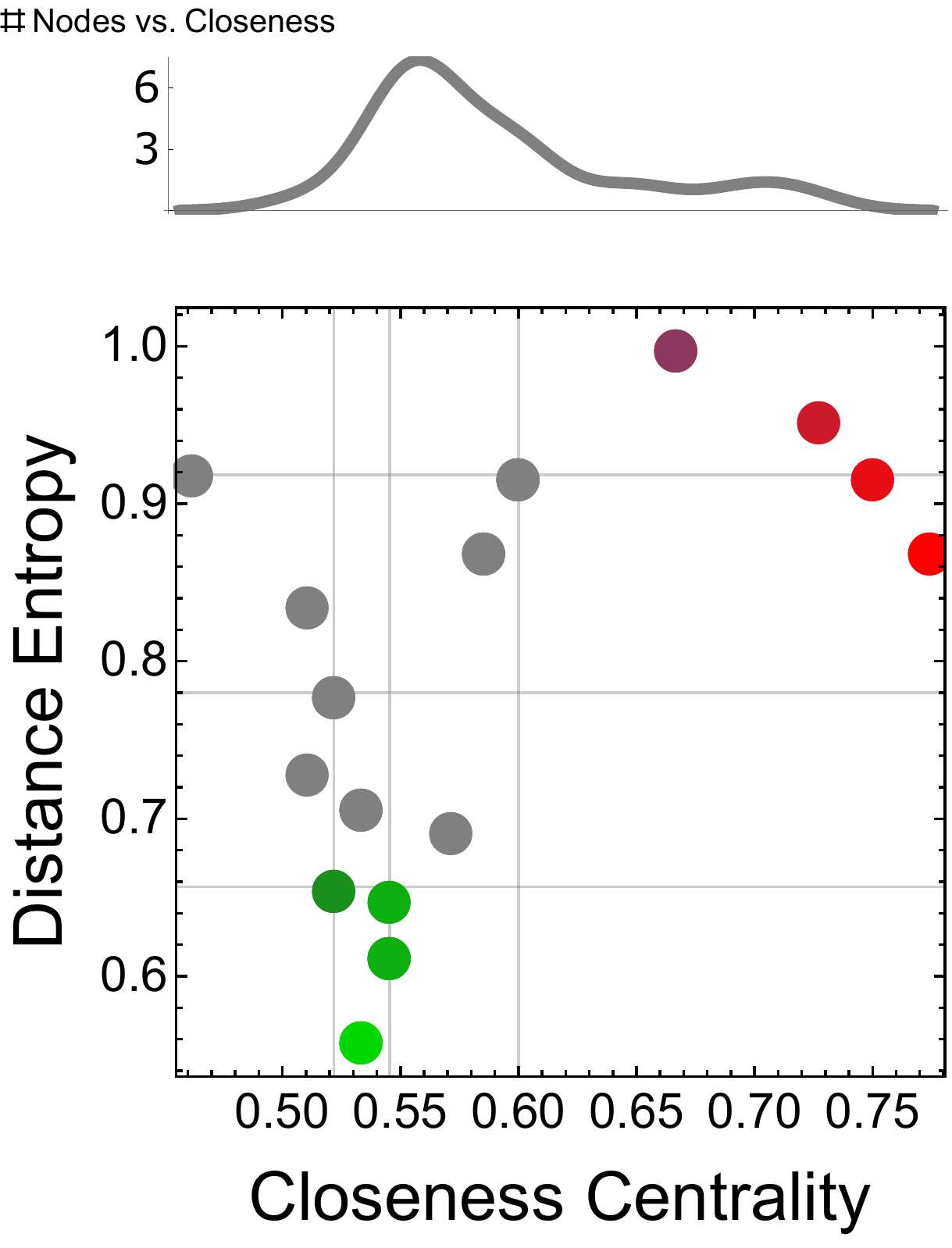}
\caption{Distance entropy provides different centrality information on nodes, compared to closeness centrality. Here we consider a BA network with $N=25$ nodes and $m=4$. (\textbf{Left}) Nodes with low distance entropy are highlighted in green. (\textbf{Middle}) Nodes with high closeness are highlighted in red. (\textbf{Right}) Cartography representing the distance entropy and closeness centrality of individual nodes in the network. Gray lines indicate quartiles. Nodes with the lowest (highest) distance entropy (closeness) are highlighted in green (red). The two sets of nodes do not overlap. Not considering distance entropy would lead to a closeness distribution reported in the top subpanel, where many nodes would end up displaying similar closeness centrality despite their different connectivity patterns, here highlighted by their distance entropy.}
\label{fig:carto}
\end{figure}

In Figure \ref{fig:carto} we show an example of distance entropy cartography for a toy model (BA network with $m=4$ and $N=25$ nodes). Sub-panel (a) highlights nodes with the lowest distance entropy (i.e. a more homogeneous distribution of network distances) while (b) highlights nodes with the highest closeness centrality (i.e. at shortest average distance in the network). Notice that nodes with higher closeness centrality tend to have also higher distance entropy, as they are closer with each other but further apart from peripheral nodes. Consequently, the set of nodes with high closeness does not overlap with the set of nodes with low distance entropy: the two metrics provide complementary information. We define a cartography by a 2D space where each node has its distance entropy and its closeness as coordinates (sub-panel (c)). Representing nodes in this 2D space is informative. In fact, most of the nodes in the network have similar closeness centrality around $c_i=0.55$ (see also sub-panel (c), top plot) but display evidently different values of distance entropy, ranging from $h=0.53$ up to $h=0.84$. Hence, considering distance entropy can allow to reduce the degeneracy observed when considering closeness only: nodes having similar closeness centrality are found, by means of the distance entropy, to be differently connected to the rest of the network.

\section{Applying distance entropy cartography to Multiplex Lexical Networks}

We apply the cartographic analysis previously introduced to multiplex lexical networks \cite{stella2018multiplex,stella2016mental}, successfully applied for modelling trends of progressive language impairments such as aphasia \cite{castro2018multiplex} but also patterns of language development such as modelling and predicting strategies of word learning in toddlers \cite{stella2017children}. When considering multiplex lexical networks and learning, we wonder if distance entropy can provide any improvement for detecting word learning strategies in toddlers.

Here, we use the same empirical networks used in \cite{stella2017children}, i.e. multi-layer edge-coloured networks where nodes represent words, there are no explicit inter-layer links and layers represent semantic relationships (e.g. "dog" and "cat" share the feature of being an animal) and phonological similarities (e.g. "bad" and "bed" differ by one phoneme only) among words. For ranking words in the order they are learned by most English toddlers between 18 and 30 months of age we use longitudinal data from the CHILDES dataset accessed through TalkBank \cite{macwhinney2007talkbank}. The longitudinal data allows to reconstruct the fraction of children producing a certain word in a given month, i.e. production probability. Within each month, words are ranked in descending order of production probability. This ranking represents a proxy for normative learning of most toddlers \cite{hills2009longitudinal,beckage2016language,stella2017children}.

Recently, different network approaches have been successfully used for predicting the acquisition of words based on their network features (e.g. word degree, closeness centrality, network gaps) \cite{hills2009longitudinal,beckage2011small,sizemore2017knowledge,stella2017children}. Here, we rank words according to the introduced cartography and then compare against the ranking of the estimated age of acquisition, in which words acquired earlier (e.g. "mommy") are ranked higher than words learned later (e.g. "picture"). The extent to which an artificial ranking $r_a$ predicts the words learned according to the normative learning ranking $r_{l}$ is measured through the word gain:

\begin{equation}
g(r_a,t)=\frac{O(r_a,r_l,t)-R(r_l,t)}{t}
\end{equation}

representing at position $t$ the fraction of words predicted as \emph{correctly learned} in $r_l$ by the network ranking $r_a$ ($O(r_a,r_l,t)$), with respect to random guessing ($R(r_l,t)$). A word gain of $20\%$ when $t=200$ words have been learned means that $r_a$ predicts as correctly learned 40 words more than random ranking. 

\begin{figure}[H]
\centering
\includegraphics[width=8cm]{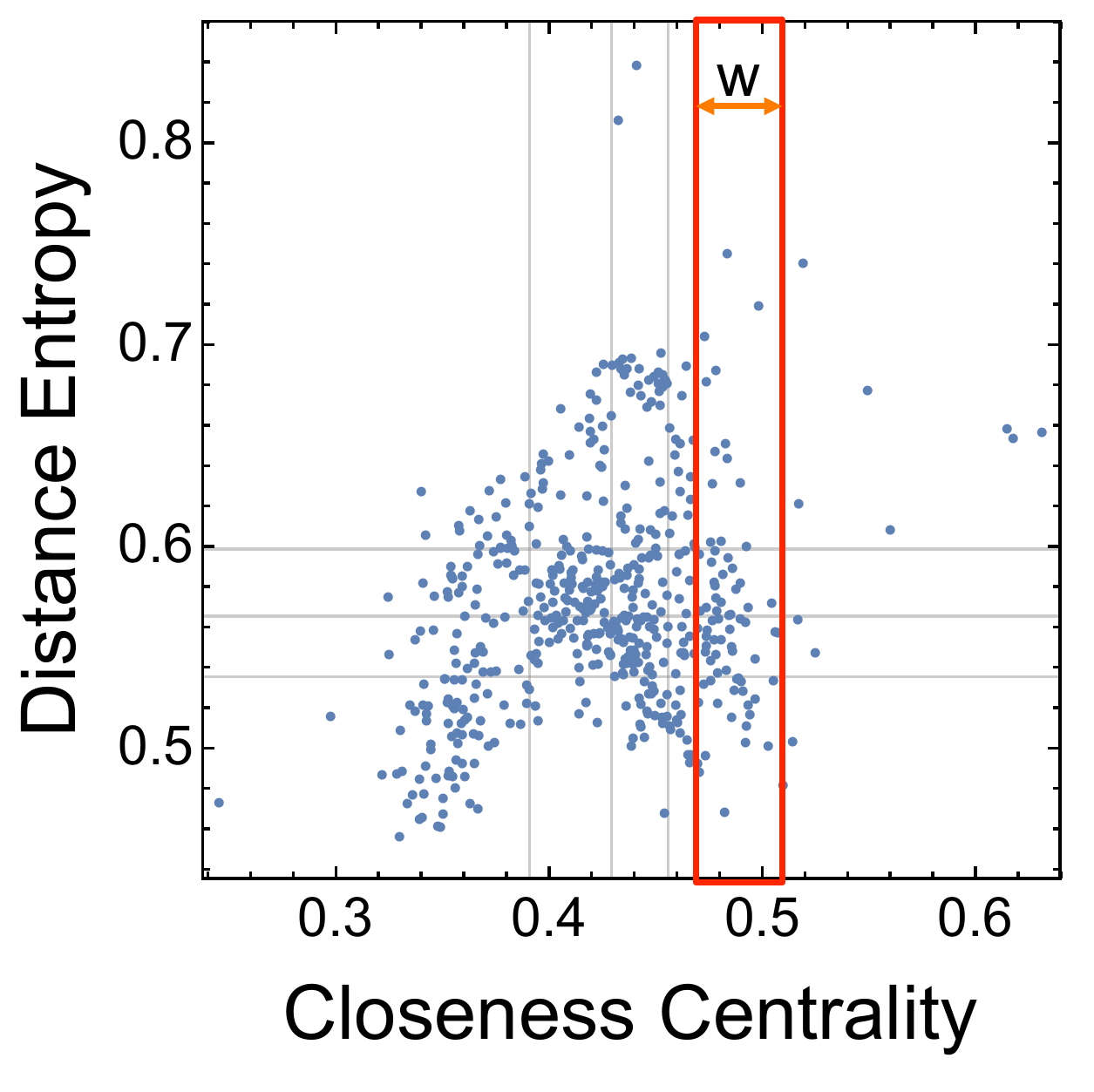}
\caption{Distance entropy cartography for the $N=529$ words in the multiplex lexical network of young toddlers. Within a window of width $w$, nodes with similar closeness centrality can have quite different distance entropies.}
\label{fig:cartchild}
\end{figure}

Multiplex closeness centrality provides a word gain higher than other measures, such as betweenness, degree and local clustering coefficient, on both single and multiplex network topologies \cite{stella2017children}. Hence, here we focus on the ranking $r_{clo}$ of descending closeness centrality and use it as a reference level to test whether enriching it with information from distance entropy can achieve higher word gains. Distance entropy is computed on the multiplex shortest paths, namely the shortest paths where links from different layers can be combined together \cite{de2013mathematical}. The resulting cartography is shown in Fig.~\ref{fig:cartchild}.

The distance entropy cartography indicates that many nodes with similar closeness centrality highly differ in their distance entropy. We quantify this notion of closeness similarity by considering nodes having closeness around $c^*$ within an interval $[c^*-w,c^*+w[$. Here $w$ represents an interval width, a tolerance parameter determining which nodes have closeness similar to $c^*$ up to a value $w$. If $w=0$ then our definition of similarity would reduce to identifying ties, i.e. considering as similar nodes having the same value of closeness $c^*$. Provided its interpretation in terms of similarity, we consider values $w \ll 1$. 

We use the cartography and closeness similarity for building artificial rankings of words based on both closeness centrality and distance entropy. Starting from the maximum value of closeness $c_{max}$ in the network, we build bins $b_i=[c_{max}-(i+1) \cdot w, c_{max}-i \cdot w [$. A ranking $r_e(w)$ is produced by ranking all nodes in bin $b_i$ in increasing order of distance entropy. Within each bin $b_i$ we rank nodes from lower to higher distance entropy because the lower $h_j$ the more a node $j$ is connected to all other nodes in the network. Hence, nodes with lower distance entropy are expected to be more central in the network. Notice that distance entropy provides different information compared to other multiplex centrality measures such as multidegree \cite{de2013mathematical} or PageRank versatility \cite{de2015ranking}, since the induced node rankings overlap with distance entropy only for $30\%$ (Kendall Tau $\tau = 0.30\%$) and $21\%$ (Kendall Tau $\tau = 0.21$) respectively. 

Notice also that when $i$ increases the average closeness of the considered nodes decreases, so $r_e(w)$ is a rank in which: (i) Highest closeness nodes are on average ranked higher, (ii) Nodes with closeness similar up to a tolerance $w$ are ranked according to their distance entropy. Hence, in $r_e(w)$ words being in the left-lower part of the entropic cartography being ranked higher, i.e. words with high closeness and low distance entropy. 

This ranking is a function of the window $w$: when $w=0$ then distance entropy has no effect and $r_e(0)$ is equivalent to ranking nodes in descending order of closeness ($r_{clo}$); when $w=1$ then all the nodes are ranked according to their distance entropy and closeness plays no role in affecting the ordering. We investigate what is the influence of $w$ in providing ranks mixing distance entropy and closeness centrality for improving prediction performances, i.e. increasing the average word gain. We focus on the early stages of cognitive development, between months 20 and 23, which are called Early Learning Stage in which multiplex closeness centrality provided the highest word gains. 

We measure increases or decreases in prediction power of which words are learned early by toddlers by considering a relative word gain improvement:

\begin{equation}
\Delta g(r_e(w),t)=\frac{g(r_e(w),t)-g(r_{clo},t)}{g(r_{clo},t)}.
\end{equation}

A relative word gain improvement $\Delta g(r_e(w),150)=0.1$ means that when $150$ have been learned, the ranking considering together closeness and distance entropy predicts as correctly learned $10\%$ more words than considering closeness only. Provided that improvements depend on the value of $w$, a scan of different values is essential. For each value of $w$, we compare the observed improvement against a distribution of random improvements obtained by fixing the same $w$ and the same bins but ranking words at random rather than according to their distance entropy. These randomised ranks represent our null models and allow one to quantify the statistical significance of word gain improvements observed when using the cartography.

\begin{table}[H]
\caption{Improvements in word gains (relative to the reference closeness case) for different values of binning width $w$. P-values are relative to the observed improvement relative to a reference distribution. Reference distributions are obtained by ranking nodes at random (rather than through distance entropy). When $w>0.05$ no improvements are obtained.}
\small 
\centering
\begin{tabular}{ccc}
\toprule
\textbf{Width w}	& \textbf{Improvement (\%)}	& \textbf{P-value}\\
\midrule
0		& 0			& 1\\
0.005		& +3.9\%			& 0.3\\
0.010		& +7.9\%			& 0.05\\
0.015		& +13.1\%			& 0.001\\
0.020		& +13.3\%			& 0.001\\
0.025		& +13.6\%			& 0.001\\
0.030		& +7.9\%			& 0.01\\
0.035		& +8.0\%			& 0.01\\
0.040		& +4.0\%			& 0.03\\
0.045		& +5.1\%			& 0.01\\
0.050		& +0.1\%			& 0.01\\
\bottomrule
\label{tabwordgain}
\end{tabular}
\end{table}

Table \ref{tabwordgain} reports the word gain improvements averaged between months 20 and 23 (the Early Learning Stage) for different values of $w$. When $0 < w \leq 0.05$ positive improvements in word gain are registered, while for $w>0.05$ only negative improvements are retrieved. The registered positive improvements are statistically significant at a 0.01 significance level when $0.015 \leq w \leq 0.025$, indicating the presence of a window where learning high closeness words with low distance entropy leads to marked improvements in predicting which words are learned by toddlers. In such cases, the average word gain achieved with the entropic cartography is $+13.3\%$.

As a comparison, ordering nodes in decreasing order of multidegree centrality within bins with $0.015 \leq w \leq 0.025$ produced on average only negative improvements of word gain ($\approx -15\%$).

These results indicate that considering distance entropy on top of the multiplex closeness centrality is beneficial in achieving better predictions of the way most English toddlers learn words. All in all, this application to real-world networks indicates that the topological information encapsulated in the distance entropy can provide additional insights in discovering and interpreting patterns of real-world complex systems.

\section{Discussion}

In this paper we introduce the distance entropy, characterising the distribution of path lengths from a given node in a network. We provide analytical results for the distance entropy of individual nodes in regular graphs and show that it is minimum when a node is adjacent to all other nodes in the network. From the analysis of the mean distance entropy in well-known network models, such as ER random graphs, BA scale-free networks and small-world networks, we observe that the creation of links either uniformly at random or through preferential attachment in general decreases the heterogeneity of path lengths, decreasing distance entropy. There is a noticeable exception in ER random graphs, where we observe a tipping point for distance entropy in increasingly denser ER random graphs. We attribute this change to the sudden emergence of short-cuts in the system. No tipping points have been detected in small-world and BA scale-free networks.

We provide evidence that distance entropy carries different topological information compared to other measures based on distance, such as closeness centrality. Consequently, we use distance entropy and closeness together in a cartography for better characterising nodes' centrality in complex networks. The additional information carried by distance entropy allows one to further distinguish nodes with equal (or very similar) closeness centralities, thanks to the fact that such nodes can be more or less heterogeneously distant from the rest of the system. 

In the current study, the concept of closeness similarity has been parametrised by means of a parameter $w$ representing the tolerance up to which two nodes are considered having similar closeness centralities. We did not fix $w$ in the current analysis in order to prevent overfitting, however its interpretation as a tolerance indicates that $w \ll 1$. Additional criteria from statistics such as using percentiles or data clustering techniques could be pursued in future work.

We use the information combined by the cartography to rank words in multiplex lexical networks \cite{stella2017children,stella2018multiplex,castro2018multiplex} and to predict the order with which toddlers learn words during cognitive development. We show that resolving the degeneracy of nodes with similar closeness centrality but (very) different distance entropy provides consistently positive improvements in predicting word learning strategies by at least $13\%$. Although this improvements might seem small two additional elements have to be considered. Firstly, multiplex closeness centrality is already an optimal measure of word prediction in the sense that it vastly outperformed single and multi-layer versions of degree, betweenness, local clustering, PageRank and eigenvector centralities in early word prediction, so that improvements to its prediction performances are remarkable. Secondly, distance entropy provides positive improvements that are not captured by other network statistics such as degree, which provides always negative word gain ($\approx -15\%$) for the same values of $w$. This result underlines the importance of considering distance entropy.

Our results provide evidence that English speaking toddlers tend to acquire mainly words with high closeness centrality and low distance entropy early on during language acquisition. These words display less heterogeneous distributions of path lengths, with smaller central moments, on the whole multiplex lexical structure and are thus easier to reach from other words in the mental lexicon. From the perspective of the cognitive sciences, the improvement in prediction of word learning indicates a cognitive advantage in learning words more central for the spread of information within the mental lexicon of word-similarities. This finding is in agreement with other independent studies indicating that closer words on linguistic networks are easier to be identified in healthy subjects \cite{goldstein2017influence} and to be produced in subjects with aphasia \cite{castro2018multiplex}.

From a network perspective, distance entropy provides different topological information compared to closeness centrality but the two measures share the same computational cost. Hence, the proposed cartography can be used for investigating also large-scale networks, providing an important tool for investigating structural patterns in real-world networks where the distances among nodes matter.

\section*{Author Contributions}

MS and MDD conceived the original study; MS designed and performed the analytical and numerical analyses. All the authors wrote the manuscript. 

\bibliographystyle{unsrt}
\bibliography{lite}

\end{document}